\documentclass[a4paper,UKenglish,cleveref, autoref, thm-restate, nolineno]{socg-lipics-v2021}

\usepackage{amsmath}
\usepackage{amsbsy}
\usepackage{amssymb}
\usepackage{algorithm}

\usepackage[table,xcdraw]{xcolor}

\usepackage{mathtools}
\usepackage{listings}
\usepackage[noend]{algpseudocode}

\algrenewcommand{\algorithmiccomment}[1]{\hfill// #1}
\newcommand{\myparagraph}[1]{\smallskip\noindent\textbf{#1}}

\newcommand{\myskip}[1]{}

\newcolumntype{P}[1]{>{\centering\arraybackslash}p{#1}}

\DeclarePairedDelimiter\floor{\lfloor}{\rfloor}
\newsavebox\CBox
\def\textBF#1{\sbox\CBox{#1}\resizebox{\wd\CBox}{\ht\CBox}{\textbf{#1}}}
\usepackage{booktabs}
\usepackage{multirow}
\usepackage{graphicx}
\newcommand{\ra}[1]{\renewcommand{\arraystretch}{#1}}
\title{GPU Computation of the Euler Characteristic Curve for Imaging Data}
\author{Fan Wang}{Stony Brook University, Stony Brook, US} {fanwang1@cs.stonybrook.edu}{}{}

\author{Hubert Wagner}{University of Florida, Gainesville, US}{hwagner@ufl.edu}{}{}

\author{Chao Chen}{Stony Brook University, Stony Brook, US}{chao.chen.1@stonybrook.edu}{}{}

\authorrunning{F. Wang, H. Wagner and C. Chen} %TODO mandatory. First: Use abbreviated first/middle names. Second (only in severe cases): Use first author plus 'et al.'

\Copyright{Fan Wang, Hubert Wagner and Chao Chen} %TODO mandatory, please use full first names. LIPIcs license is "CC-BY";  http://creativecommons.org/licenses/by/3.0/

\ccsdesc[100]{F.2.2 Nonnumerical Algorithms and Problems}
\ccsdesc[100]{Theory of computation~Computational geometry}
\ccsdesc[100]{Mathematics of computing~Combinatorial algorithms}
\keywords{topological data analysis, Euler characteristic, Euler characteristic curve, Betti curve, persistent homology, algorithms, parallel programming, algorithm engineering, GPU programming, imaging data} %TODO mandatory; please add comma-separated list of keywords

\category{} %optional, e.g. invited paper

\relatedversiondetails{Full Version}{https://arxiv.org/abs/2203.09087}
\supplementdetails{Source Code}{https://github.com/TopoXLab/GPU_ECC_SoCG2022}
\funding{This work was partially supported by grants NSF IIS-1909038 and CCF-1855760.}%optional, to capture a funding statement, which applies to all authors. Please enter author specific funding statements as fifth argument of the \author macro.

\EventEditors{Xavier Goaoc and Michael Kerber}
\EventNoEds{2}
\EventLongTitle{38th International Symposium on Computational Geometry (SoCG 2022)}
\EventShortTitle{SoCG 2022}
\EventAcronym{SoCG}
\EventYear{2022}
\EventDate{June 7--10, 2022}
\EventLocation{Berlin, Germany}
\EventLogo{figs/socg-logo}
\SeriesVolume{224}
\ArticleNo{63}
\begin{document}

\maketitle

\begin{abstract}
Persistent homology is perhaps the most popular and useful tool offered by topological data analysis -- with point-cloud data being the most common setup. Its older cousin, the Euler characteristic curve (ECC) is less expressive -- but far easier to compute. It is particularly suitable for analyzing imaging data, and is commonly used in fields ranging from astrophysics to biomedical image analysis. These fields are embracing GPU computations to handle increasingly large datasets. 

We therefore propose an optimized GPU implementation of ECC computation for 2D and 3D grayscale images. The goal of this paper is twofold. First, we offer a practical tool, illustrating its performance with thorough experimentation -- but also explain its inherent shortcomings. Second, this simple algorithm serves as a perfect backdrop for highlighting basic GPU programming techniques that make our implementation so efficient -- and some common pitfalls we avoided. This is intended as a step towards a wider usage of GPU programming in computational geometry and topology software. We find this is particularly important as geometric and topological tools are used in conjunction with modern, GPU-accelerated machine learning frameworks.
\end{abstract}

\section{Introduction}
\label{sec:intro}
Describing the shape of data is the tenet of topological data analysis (TDA) -- and at its heart lies the idea of studying data across scales. Instead of characterizing the shape at a fixed scale -- we measure its evolution. A filtration encodes this evolution and thus becomes an object of primary interest. Depending on the type of data, an appropriate filtration is used: Alpha-shape filtration for point-cloud data embedded in three dimensional space; Vietoris--Rips filtration for high dimensional metric data expressed by pairwise distances; cubical filtration for two- or three-dimensional grayscale imaging data. This paper focuses on imaging data, in which TDA methods have shown promise in recent years \cite{WangKLPC21, topogan, optimalcyc, hxlseg}.

Persistent homology is perhaps the most powerful topological descriptor applied to such filtrations -- and it proves especially useful in conjunction with modern deep learning (DL) methods. However, integration of persistent homology with DL methods remains far from seamless -- despite significant progress, computing persistent homology takes significant amount of time and resources for practical datasets. This is in contrast with modern learning pipelines which often employ simple, highly optimized computations. In particular, many neural network architectures are realized fully on graphical processing units (GPUs) attaining massively parallel processing; the same applies to modern large-scale simulations. Existing software for persistent homology is not at this level of advancement, at least not for imaging data. 
We mention that the recent GPU implementation by Zhang et al.~\cite{zhang2020gpu} is in the context of Vietoris--Rips filtrations coming from point-cloud data and cannot handle imaging data.

% Case in point: state-of-the-art package CubicalRipser compute the persistence diagram of an $512^3$ image in 6 minutes~\cite{kaji2020cubical}. This is a big achievement brought about by two decades of improvements. However, we estimate that a truly seamless integration with ML software would require doing this computation 30 times per second \cc{Where does this number come from?}. This requires further speedup exceeding the factor of 10000. \cc{The last sentence is a bit hard to understand.}

In view of the above, we turn our attention to a simpler -- but still expressive -- topological descriptor, namely the Euler characteristic curve (ECC). %; we also focus entirely on filtrations derived from imaging data. 
ECC has an excellent track record in providing relevant topological information in various imaging applications~\cite{primoz,barley,crawford2020predicting} -- we elaborate on this in Section~\ref{sec:related}. More importantly, we demonstrate that we can compute ECC at extremely fast speed -- for example we can process a 3D image of size $512^3$ 30 times per second. We also managed to implement a streaming strategy, which allows us to handle huge images of $4096^3$ and beyond -- despite the limited GPU memory. The above points imply that a truly seamless integration with modern image processing pipelines is achievable. Overall, we hope to impact the following field. 
% ML pipelines and physical simulations is achievable. 

% As we hinted in Section~\ref{sec:related}, the ECC has proven to be useful in different fields. Our design choices are heavily affected by the scenarios in which GPU computations of ECC promise to be useful. We mention two such scenarios here.

\myparagraph{Machine learning.} We are particularly interested in incorporating ECC computation into machine learning frameworks, e.g., convolutional neural networks (CNNs) for computer vision~\cite{alexnet}, biomedical image processing~\cite{cnns_in_medical} or computational astrophysics~\cite{cnn_in_astro}. In these contexts, ECC can be used as topological features for prediction models.
% CNNs methods were first introduced in the 1980s -- but large-scale applications became feasible only thanks to modern GPUs. An interesting study~\cite{li2016performance} of implementations of CNNs shows that relying on a CPU implementation for any image operation is likely to cause a significant performance bottleneck -- even if this implementation is highly optimized. The primary reason quoted in~\cite{li2016performance} is the large cost of transferring data between the GPU memory and the main memory (RAM). 
% This was our primary motivation to work on the GPU implementation -- knowing that even a highly optimized, parallel CPU implementation of ECC would be insufficient in this context. In this scenario ECC computation becomes par of the pipeline -- and our focus is on quick handling of an images already residing in GPU memory.

\myskip{
\myparagraph{Physical and astrophysical simulations.} 
% In the second scenario we focus on large images -- or large volumetric data in general. The simplest setting is handling a single large file. Our streaming GPU implementation can handle files of arbitrary size, we tested volumes up to $4096^3$ voxels. However our real motivation comes from physical and astrophysical simulations. 
Increasingly frequent simulations in these domains are implemented on GPUs and result in large 3D volumes~\cite{big_astro, vol_simulation}. We are motivated by the prospect of performing ECC computations \emph{in vivo}, namely during the simulation, exploiting the fact that the simulated volume resides in GPU memory.}

\myparagraph{Contributions.} The main technical contribution of this paper is a streaming GPU implementation of ECC computation for imaging data. While the underlying algorithm is very simple, our contribution lies in the implementation carefully tuned to modern GPUs. In particular, when adapting computation into massive parallelism, we need to carefully design the implementation so that the limited GPU memory resources can be exploited in the most efficient manner. %Furthermore, data transfer latency (i.e., loading data into RAM) is a major bottleneck, we design the algorithm to best exploit the asynchronous behavior of a GPU and GPU/CPU parallelism so that the data transfer latency is not a bottleneck.
% We will carefully explain our implementational choices -- without assuming prior familiarity with GPU concepts. 

\myskip{It is also our hope that by demystifying some of these low-level details, we are making a step towards a wider adoption of GPUs in computational geometry and topology software. In particular, we do believe that similar techniques could be used to provide further improvement in persistence computations in the context of image analysis.}

\myskip{
\myparagraph{Outline.} This paper is structured as follows. In Section~\ref{sec:background}, we briefly define ECC and related concepts.
In Section~\ref{sec:related}, we mention existing applications and implementations of ECC. In Section~\ref{sec:algorithm} we explain our GPU implementation; some more advanced optimizations are covered in the Appendix in Section~\ref{sec:optimization}. In Section~\ref{sec:experiment} we provide thorough experiments and discuss them vis-a-vis our goals. We summarize the paper in the last section.}

\section{Background}
\label{sec:background}
%In this section we briefly cover the basics needed for topological treatment of imaging data. %Then we discuss basic concepts related to GPU programming. 

\subsection{Images as cubical filtrations}
\label{cubical_filt}
The input to our algorithm is a $d$-dimensional grayscale image, by which we simply mean a $d$-dimensional array of real values. Individual elements are called pixels (in 2D) and voxels (in 3D and above), and we collectively call them voxels. One common operation is \emph{thresholding}, which selects the subset of voxels not exceeding a certain threshold $t$. To talk about the topology of a sequence of thresholdings, we impose more structures on the data.

\begin{algorithm}
\caption{Sequential computation of the VCEC}
\label{alg:cpualg}
\begin{algorithmic}[1]

\Require $\boldsymbol{I}$: an input image
\Ensure $\boldsymbol{VCEC}$: the vector of changes in the Euler characteristic.
\State initialize $\boldsymbol{VCEC}$ as an empty array
\ForAll{voxels $v$ in $\boldsymbol{I}$}
\ForAll{faces $c$ of $v$}
\label{alg:from}
\If {$c$ was introduced by $v$} 
\State $\boldsymbol{VCEC}[\text{value of } v \text{ in } I] \leftarrow \boldsymbol{VCEC}[\text{value of } v \text{ in } I] + (-1)^{\text{dimension of face }  c}$ 
\EndIf
\label{alg:to}
%\algorithmiccomment{dim(c): dimension of face $c$}
\EndFor
\EndFor 
\end{algorithmic}
\end{algorithm}

To this end we follow~\cite{kmm}. First, we define an elementary interval as either $[k,k+1]$, or a degenerate interval $[k,k]$, for an integer $k$. An elementary (cubical) cell is a product of $d$ elementary intervals, and its dimension is the number of non-degenerate intervals entering its product. This way we can talk about vertices, edges, squares, cubes etc as cells of dimension $0,1,2,3$ etc. We say that cell $a$ is a face of cell $b$ iff $a \subset b$, or a coface if $b \subset a$. Now, we associate the input values with the top dimensional cells, which we call \emph{voxels}. Finally we extend the values from voxels to all lower dimensional cells: each cell inherits the minimum value of its top-dimensional cofaces. The thresholding of the image at value $t$ is now a \emph{cubical complex}, $K_{\leq t}$; the nested sequence of these complexes form a \emph{cubical filtration} indexed by $t$.

\subsection{The Euler characteristic curve}
With the above setup, we can define the Euler characteristic curve of a cubical filtration as the sequence
\begin{equation}
    ECC_i = \chi({K_{\leq t_i}}) = \sum_{j}(-1)^j c_j({K_{\leq t_i}}) = \sum_{j}(-1)^j \beta_j(K_{\leq t_i})
\end{equation}
where $t_i$ is the i-th smallest grayscale value in the image, $c_j(.)$ counts the $j$-dimensional cells, and $\beta_j(.)$ is the $j$-dimensional Betti number. The last equality comes from the Euler--Poincare formula and ties ECC with the topology of the space.

\begin{figure}[ht!]
    \centering
    \includegraphics[width=0.80\textwidth]{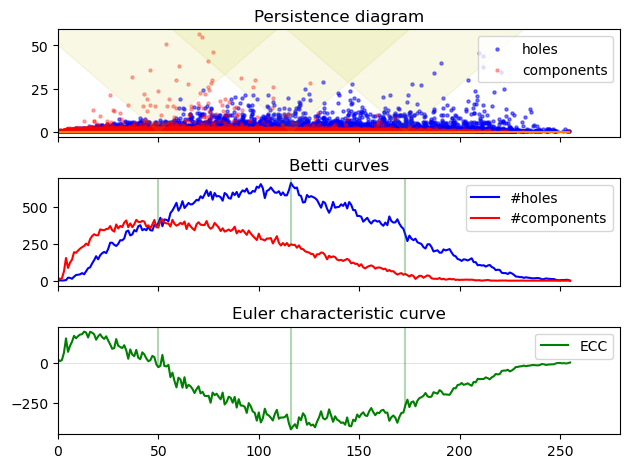}
    \centering
    \includegraphics[]{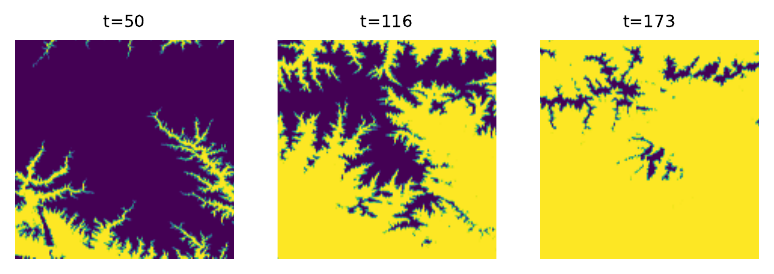}

    \caption{({\bf bottom}) Input image at three thresholds. Their grayscale values correspond to terrain elevations. ({\bf top}) The three plots share the x-axis which represents the thresholds of the input image. The topmost plot shows the persistence diagram (rotated for clarity). For each threshold, it marks the lifetime of topological features: connected components in red, and holes in blue. The three highlighted areas show the features alive at the three corresponding thresholds, which are visualized as the Betti curves below. The ECC is the pointwise difference between the curves. This example highlights the main downside of ECC: its reliance on counts of topological features, while persistence also distinguishes their prominence.}
    \label{fig:pers_to_euler}
\end{figure}

We only mention that the Betti numbers are the ranks of the \emph{cubical homology groups}~\cite{kmm} of the cubical complex $K_{\leq t}$. For three dimensional complexes, the Betti numbers count the number of connected components, tunnels and voids in an object. Therefore, the ECC mixes up the numbers of topological features at each threshold. We can also define \emph{persistent homology}~\cite{edelsbrunner2000topological} in this setup~\cite{wagner2012efficient}. Fig.~\ref{fig:pers_to_euler} illustrates the relationship between persistent homology, the Betti curves and the ECC.

\subsection{ECC computation}
A naive algorithm for ECC explicitly computes the Euler characteristic (EC) at each threshold. This results in time complexity of $O(mn)$, where $m$ is the number of unique values in the image, and $n$ the number of voxels. We assume the dimension of the image is  a constant.

An algorithm by Snidaro and Foresti~\cite{rtth} was the first offering $O(n)$ complexity, but is quite complicated and hard to generalize beyond 2D. Our approach is  based on a much simpler algorithm~\cite{chunkyeuler}, which interprets an image as a cubical filtration. 

\myparagraph{Tracking the VCEC.} The main idea is to compute the "Vector of Changes in the Euler Characteristic" (VCEC), namely a sequence of length $m$ such that $VCEC_0 = ECC_0$ and $VCEC_i = ECC_i-ECC_{i-1}$, for $0 < i < m$. Since $ECC_i = \sum_{j=0}^{i} VCEC_j$ by construction, we compute in time $O(m)$ with basic dynamic programming -- although since $m$ is small, a $log(m)$ parallel algorithm is a practical alternative on GPUs~\cite{sengupta2008efficient}.

%In other words, $VCEC_i$ computes the Euler characteristic of the cells with value $t_i$. We note that this subset of cells forms a chain complex (so we can talk about homology groups), however generally it does not  form  a cubical complex (so the intuitive interpretation generally does not extend to  the ranks of these homology groups; we could interpret them in the language of relative homology, but this is not necessary here).

\myparagraph{Faces introduced by a voxel.} We say that a face is \emph{introduced by a voxel}, if this voxel has the smallest value among all voxels containing the given face (in other words: if the face inherits the value from this voxel). One caveat is that ties have to be broken in a consistent manner: if two voxels have the same value then we prefer the one with a lexicographically lower position in the input array. Later we show that this turns in a fast computation, without the need to explicitly compare the indices.

\myparagraph{Sequential algorithm.} With this we can sketch a simple sequential algorithm for the computation of VCEC for an image (see Algorithm~\ref{alg:cpualg}). Note that there is no need to explicitly store any information of the lower dimensional cells. This algorithm will be a basis for our GPU algorithm. 

%\includegraphics[]{figs/desc.pdf}

% \subsection{GPU computations}
% We  outline the most important aspects related to graphical processing units, commonly called GPUs. More specific techniques are discussed along with our implementation in Section~\ref{sec:algorithm}.

% This type of hardware was originally introduced to accelerate 2D graphics, and then rendering of 3D scenes. Since each pixel of the resulting 2D image requires similar but independent computations, these computations are parallizable. Currently, GPUs are commonly used for general purpose processing. Not all problems are a good fit for this kind of computations -- but ECC computation is.

% We focus on the nVidia GPU hardware, and their CUDA programming framework, due to their popularity. In this programming model the programmer provides a \emph{kernel}, which is a function that is run in parallel on a large number of threads. For example the nVidia RTX 2070 GPU we use in our experiments offers 2304 hardware threads of computation. This GPU realizes the Turing architecture, and the information contained in this paper are often specific to this modern architecture. Older GPUs made different architectural choices, which often required specific programming techniques.% In general these architectural changes slowly make GPU programming more programmer-friendly -- which we also exploit in our implementation.

\section{Related work on ECC: applications and computations}
\label{sec:related}
Due to their simplicity, both the Euler characteristic curve (ECC) and the Euler characteristic (EC) find usage in many fields -- especially the ones related to imaging. A great introduction to the topic is the review paper by Worsley \cite{worsley1996geometry} from which we sample some of the applications below. We then discuss the related work on the computational side.

\begin{figure}[ht!]
    \centering
    \includegraphics[width=0.99\textwidth]{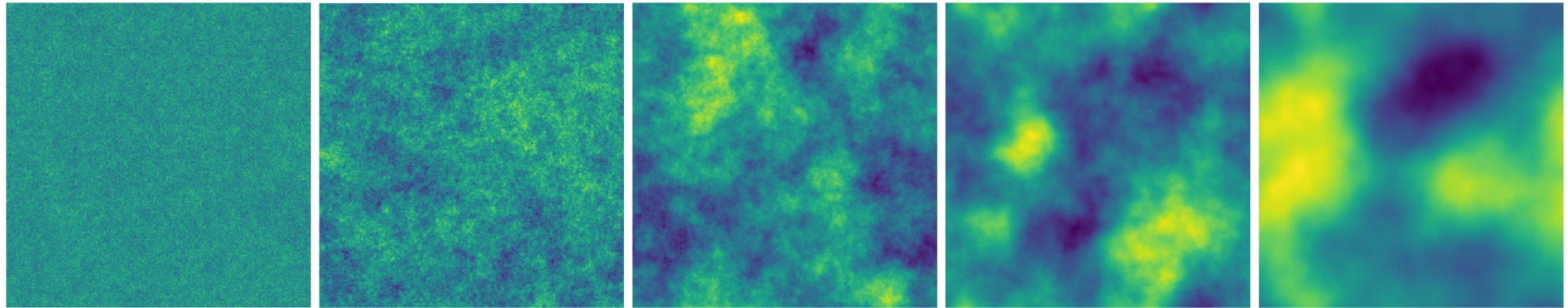}
    \caption{Visualizations of Gaussian random fields generated with different levels of smoothness.}
    \label{fig:grfs}
\end{figure}

\myparagraph{Applications.}
Ideas related to the EC were present in astrophysics already in 1970s (ECC is called the \emph{genus curve}). They were formalized in 1986 by Gott and others~\cite{gott1986sponge} in the study of the sponge-like topology of the large-scale structures in the universe; later ECC became an important tool in the study of the imaging data describing the cosmic microwave background (CMB) radiation~\cite{novikov}. This is closely related to earlier work on the topology of Gaussian random fields (GRFs) by Adler and Hasofer~\cite{robert_random_fields} -- GRFs are used to model the CMB. See Fig.~\ref{fig:grfs} for images of GRFs. Ideas related to the EC were popular in the field of bone morphometry. They were formalized mathematically in 1993~\cite{odgaard1993quantification}; there EC was used to characterize the trabecular structures in bones -- particularly to compute the first Betti number (called the \emph{connectivity} in this field). EC is a common tool in \emph{morphological image processing}~\cite{horn1986robot}; it is widely used to characterize the shape of thresholded (binary) images under the name \emph{Euler number}; later it was computed at all thresholds of a grayscale image -- this is ECC hiding under the name \emph{stable Euler number}~\cite{rtth}. In particular the zero-crossing of the ECC is used to select a segmentation threshold; see Fig.~\ref{fig:pers_to_euler} for a rudimentary example showing that the riverbeds in a terrain are clearly highlighted at this threshold.

\myparagraph{ECC in TDA.} Many of the above applications are close to the way topological descriptors are used today in topological data analysis (TDA), although persistent homology is a much more popular choice. Still, there is a number of recent TDA studies using EC and ECC. Bobrowski and Skraba~\cite{primoz} demonstarate that ECC is surprisingly powerful in analyzing the percolation threshold in random cubical filtrations (and other random models). Crawford and collaborators propose~\cite{crawford2020predicting} as a novel statistic based on EC; it proves useful in predicting clinical outcomes of brain cancer based on brain imaging data. Amezquita and collaborators~\cite{barley} analyzed the shape of barley using an image transform based on EC.

\myparagraph{Computations.} However, the employed algorithmic techniques were different from the simple setup outlined in the previous section. Instead, the computations often exploited the connection of EC with differential geometry. In particular, an explicit, efficient algorithm for EC of 3D voxel data was presented in~\cite{gott1986sponge}. Its efficiency stems from precomputed tables of voxel neighbourhood. Our approach is different and based on the mathematical-algorithmic setup of cubical homology. This direction emerged in the 1990s in the work of Kaczynski, Mischaikow and Mrozek. 
Originating in the context of computational dynamics, it evolved in a more general framework described in their book~\cite{kmm}. The first efficient, general-dimension algorithm for EC of binary images uses this setup. The algorithm is due to Ziou and Allili~\cite{Ziou2002} in 2001. The idea is to view a binary image as a cubical complex and compactly encode this information. Compared to existing algorithms, this approach is simple, efficient, and it works in arbitrary dimension. One drawback is the memory overhead related to storing the cubical complex.

%Our approach is different and based on the mathematical-algorithmic setup of cubical homology. This direction emerged in the 1990s in the work of Kaczynski, Mischaikow and Mrozek.

The first efficient algorithm for ECC computation is presented by Snidaro and Foresti~\cite{rtth} in 2003. It  focused on 2D images. The first efficient algorithm for ECC for 3D images -- which also works in arbitrary dimension -- is due to Heiss and Wagner~\cite{chunkyeuler} in 2017. This approach extends the idea of Ziou and Allili to cubical filtrations (i.e. from binary to grayscale data). It also offers improvements: the cubical complex is not stored explicitly; computations are done in parallel; the image is streamed into memory in small chunks so that images of arbitrary size can be handled. Our GPU implementation is based on this approach.

% However, the employed algorithmic techniques were different from the simple setup outlined in the previous section. Instead, the computations often exploited the connection of EC with differential geometry. In particular, an explicit, efficient algorithm for EC of 3D voxel data~\cite{gott}. Its efficiency stems from precomputed tables of voxel neighbourhood.

\begin{figure*}[ht!]
\centering
\includegraphics[width=.9\textwidth]{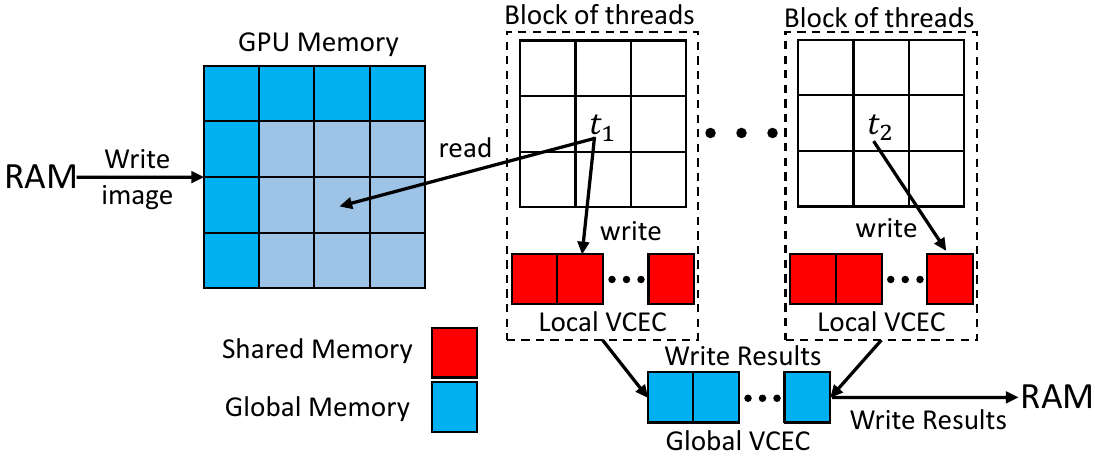}
\caption[ph]{The image is first copied from RAM to GPU's global memory. Each block of threads is responsible for a patch of the image. Each thread in the block needs to access a voxel and its eight neighbours (all marked in lighter blue). When the block is done, the block's local result is added to the global result. The final result for the entire image is transferred to RAM.}
\label{fig:gpualgo}
\end{figure*}

\section{GPU implementation}
\label{sec:algorithm}
Our GPU implementation is illustrated in Algorithm~\ref{alg:gpualg}. The listed code is slightly simplified for readability and covers only the case of 2-dimensional images with grayscale levels ranging from 0 to 255. After we cover the overall structure of computations, we explain the implementation in details.

\subsection{Challenges}
Our GPU implementation tackles three main challenges: (1) We needed to adapt the CPU algorithm to the GPU setting to fully exploit its massive parallelism. The main challenge was that instead a dozen of threads we had to manage thousands of threads working in parallel -- which required us to structure the computations differently. Further, unlike the CPU version, which employed a simple lock-free scheme, we needed to explicitly deal with race conditions and other issues related to synchronization. Apart from ensuring correctness, we had to experiment with different synchronization granularities to achieve optimized performance.

(2) Efficient use of GPU's memory hierarchy and its limited resources. GPU is notorious for its complicated memory hierarchy and limited memory resources. Unlike CPU programming these have to be explicitly incorporated into algorithmic design. We managed to craft an efficient multi-level caching hierarchy, taking into account the access patterns characteristic of working with cubical complexes. With careful analysis of access probabilities, we managed to ensure that only a single voxel (and not 9 or 27) per thread is fetched from main memory.   (3) Many of the technicalities are not visible when analyzing the GPU kernel. One particular technical difficulty was achieving streaming operation without affecting the performance. This enabled us to handle inputs of virtually unlimited size, despite limited GPU memory. We also organized the streaming processing in a pipeline which allows  to overlap the computations with memory transfers.

\subsection{Structure of the computations} The C++ implementation shown in Algorithm~\ref{alg:gpualg} defines a \emph{compute kernel} which is the computation realized by a thread. Thread-centric view is assumed hereafter, and we will talk about \emph{the thread} remembering that the same computations are done concurrently by many threads. %In this sense the perspective is similar to designing an algorithm for a PRAM machine -- the main difference is that hardware considerations come to a foreground.

\myparagraph{Single thread.} Each thread handles a single voxel, namely, realizes lines \ref{alg:from}--\ref{alg:to} of Algorithm~\ref{alg:cpualg}. In other words, each thread iterates over the faces of a given voxel, decides which of them are introduced by this voxel, and updates the VCEC vector at the value of the voxel.

\myparagraph{Blocks of threads.} Threads are grouped into \emph{blocks}, and we can imagine that the image is decomposed into rectangular patches. Many such patches are processed concurrently -- although not necessarily in parallel. See Fig.~\ref{fig:gpualgo} for an overview.

\subsection{Optimizations}
\label{optimizations_}
Computations structured this way perfectly fit the GPU pipeline -- however, using the potential of the hardware requires careful memory management. Specifically, GPUs have a hierarchy of memory types with different sizes and performance characteristics. Unlike CPU programming, the programmer must make thoughtful use of these various kinds of memories.

Below we explain some of our implementation choices, and mention common pitfalls. To illustrate these issues, we start from a hypothetical naive implementation directly implementing Algorithm~\ref{alg:cpualg} in GPU. We then improve it step by step, arriving at the implementation listed in Algorithm~\ref{alg:gpualg}.

\myparagraph{Location of input image.} The image data is initially copied from main memory (RAM) to the GPU's \emph{global memory} -- this is the only type of GPU memory large enough to store an image of a reasonable size. Note that large images may exceed the size of the global memory. For now we assume that the image fits in global memory -- we solve this issue in Section~\ref{sec:streaming}.

\myparagraph{Race conditions.} We store the VCEC in global memory, simply as an array of 256 integers. However, since multiple threads will update the same memory location, we need to be wary of \emph{race conditions}. Modern GPUs offer efficient implementation of \emph{atomic operations}, including \emph{atomicAdd}, which ensures that all the updates to VCEC will be correctly recorded. However, all updates issued simultanously on a single memory location will be \emph{serialized} -- which means we lose the main advantage of the GPU hardware, namely, massive parallelism.

\myparagraph{Using registers.} We can mitigate the above problem by accumulating the contribution of the given voxel in a \emph{register}. Registers provide the fastest type of GPU memory. Additionally, they are local to each thread, which means that we do not need to worry about race conditions when updating the values stored in them. We still need to update the global VCEC using atomicAdd, but the number of updates is now one per thread (instead of 9 in 2D or 27 in 3D).

\myparagraph{Shared memory.} The above is an improvement but still far from ideal. The next step is to use \emph{shared memory}. This is another type of low-latency memory offered by GPUs, although slower than registers. It is shared between all threads in a given block. So instead of updating the global VCEC, each thread updates the local VCEC of its own block. This local VCEC is simply an array of 256 integers, as declared in line \ref{gpu:shared}.

We still need to use atomicAdd to avoid race conditions -- but now the collision probability is lower, since only threads belonging to a single block can access a given shared memory location. And since the size of the block is configurable, we can find the size which yields good performance. This optimization has two additional advantages: shared memory has significantly lower latency than global memory; in modern GPUs atomic operations on shared memory are significantly more efficient than on global memory. This is not true on older GPUs, and using them would require a more elaborate way of merging the results.

\begin{figure*}[ht!]
\centering
\includegraphics[width=.99\textwidth]{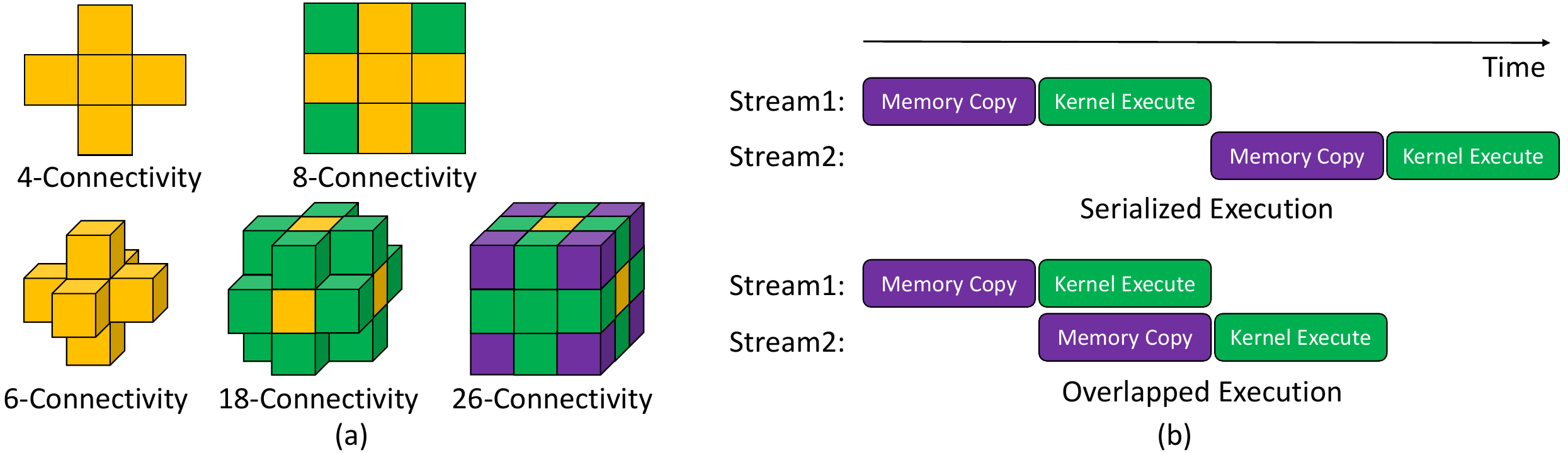}
\caption[ph]{(a) Various kinds of connectivity relations between voxels. (b) An illustration showing that overlapping the computations and memory transfer can reduce the overall execution time.}
\label{fig:sync_connect}
\end{figure*}

\myparagraph{Parallel initialization and finalization.} The shared memory needs to be initialized, and its final content needs to be added to the global VCEC vector. We perform both steps in parallel: a single thread in the block is responsible for one location in the shared-memory array. In line~\ref{gpu:zero} the thread sets a specific location to zero -- or does nothing. Similarly, in line~\ref{gpu:atomic} the thread issues an atomicAdd, which updates the global result with the local one. Note that in these two cases the index does not depend on the value of the voxel assigned to the thread -- we simply compute the unique number of the thread within its block; see line~\ref{gpu:index}. We use it to index the shared array. 

\myparagraph{Block-level synchronization.} Since the threads in the block are not guaranteed to run in parallel, we need to synchronize them -- otherwise they could start working on uninitialized memory, or update the global result using unfinished local results. Proper synchronization is insured by placing a \emph{block-level synchronizing barrier} in lines \ref{gpu:sync} and \ref{gpu:sync2}.

\myparagraph{Accessing neighbors.} To decide which cells are introduced by a given voxel, we compare the value of the voxel with its neighbours (with careful tie-breaking). Specifically, we access the 8-connectivity neighbors of a given voxel (and 26 in 3D); see Fig.~\ref{fig:sync_connect}(a) for an illustration. 

\myparagraph{Texture cache.} Accessing these values is another source of inefficiency, linked with the high latency of the global memory. We mitigate this by using a specialized caching mechanism, called \emph{texture cache}. It is often referred to as \emph{texture memory}, which is misleading since modern GPUs realize this as a caching layer on top of data residing in global memory. When the value of a voxel is requested from global memory using this mechanism, the neighbours of the accessed voxel are automatically cached in GPU's specialized low-latency memory. It is as fast as shared memory. Line \ref{gpu:reg} shows how a voxel value is requested via the texture cache. This is a read-only cache, which suits our algorithm well, since we are not modifying the image.

The texture cache is optimized exactly for the \emph{spatial access locality} displayed by ECC computations. Also, since neighboring voxels are generally processed in parallel, it is likely that the neighbors' values already reside in the fast cache. Overall, we can expect that on average only a single uncached global memory access will be required per thread -- but there is no guarantee due to the limited size of the cache.

\myparagraph{The danger of register spilling.} We store the frequently used voxel values in \emph{registers}. In the case of 2D inputs, the 4-connectivity neighbors (marked in yellow in Fig. \ref{fig:sync_connect}(a)) are involved in 3 different comparison operations and therefore we cache them in registers. The 8-connectivity voxels (marked in green) are read once and used once, so we save registers and rely on the aforementioned texture cache. Similarly for 3D inputs, only the 6-connectivity voxels (used 9 times) and 18-connectivity voxels (used 3 times) are stored in registers. To use the registers, we unroll the loop, namely, replace it with a series of statements.

\begin{algorithm}[H]
\caption{Implementation of the VCEC on GPU for a 2D image}
\label{alg:gpualg}
\begin{lstlisting}[language=C++, escapechar = |]
__constant__ int image_width, image_height; |\label{gpu:const}|
const int num_bins = 256;

__global__ void vcec_kernel(cudaTextureObject_t voxels, int* vcec_global)
{
    |\label{gpu:shared}|__shared__ int vcec_local[num_bins]; 
    |\label{gpu:index}|const int thread_number = blockDim.x * threadIdx.y + threadIdx.x;

    if (thread_number < num_bins)
        vcec_local[thread_number] = 0; |\label{gpu:zero}|
    __syncthreads(); |\label{gpu:sync}|

    const int ix = blockDim.x * blockIdx.x + threadIdx.x + 1;
    const int iy = blockDim.y * blockIdx.y + threadIdx.y + 1;
    if (ix >= image_width + 1 || iy >= image_height + 1) return;

    int change = 1; |\label{gpu:reg}|
    int c = tex2D<float>(voxels, ix, iy); |\label{gpu:tex}|
    int t = tex2D<float>(voxels, ix, iy - 1);
    int b = tex2D<float>(voxels, ix, iy + 1);
    int l = tex2D<float>(voxels, ix - 1, iy);
    int r = tex2D<float>(voxels, ix + 1, iy);

    // Vertices 
    |\label{gpu:verts_begin}|change+=(c < l && c < t && c < tex2D<float>(voxels, ix - 1, iy - 1));
    change+=(c < t && c <= r && c < tex2D<float>(voxels, ix + 1, iy - 1));
    change+=(c < l && c <= b && c <= tex2D<float>(voxels, ix - 1, iy + 1));
    change+=(c <= b && c <= r && c <= tex2D<float>(voxels, ix + 1, iy + 1)); |\label{gpu:verts_end}|
    // Edges
    change -= ((c < t) + (c < l) + (c <= r) + (c <= b));|\label{gpu:edges}|

    atomicAdd(&vcec_local[c], change); |\label{gpu:atomic}|
    __syncthreads(); |\label{gpu:sync2}|
    if (thread_number < num_bins)
        atomicAdd(&vcec_global[thread_number], vcec_local[thread_number]); |\label{gpu:update_global}|
}
\end{lstlisting}
\end{algorithm}

It may seem like a good idea to cache everything in registers. This is especially misleading since defining register variables is syntactically the same as defining stack-allocated variables in C++ CPU programming; see line~\ref{gpu:reg} for an example. We are careful with register allocation -- one common pitfall is \emph{register spilling}. One danger stems from the fact that the number of registers per block is limited by hardware -- but the number of threads in the block is configurable. So requesting a block of a certain size may cause the number of required blocks to exceed the availability. In this case the values are -- silently! -- stored in what is called \emph{local memory} -- which is a misnomer because the values are physically placed in global memory. So instead of using the fastest memory, the slowest one is used. This simple mistake can cause a performance hit of two orders of magnitude.

\myparagraph{Branching.} GPU performance can be significantly penalized by branching and loops. In general, GPU programs allow for considerable flexibility -- but they operate most efficiently as SIMD (single instruction multiple data) units. In other words, kernels whose flow of execution does not depend on the input data are preferred.

\myparagraph{Warps of threads.} The above is related to how the threads are scheduled by GPU. Namely, the threads within each block are additionally grouped into \emph{warps} of 32 threads. Any branch (i.e. if statement) splits the execution of the entire warp into two divergent paths. This is often called \emph{intra-warp branching}. These two paths generally cannot be executed in parallel, instead they are serialized. So a single thread can cause all the remaining threads in its warp to remain idle, limiting parallel execution. We avoid branching in several ways: we surround the data with a collar of voxels with infinite value, to avoid divergent branches due to boundary conditions. We also update the \emph{change} variable without a branching statement, see e.g. line~\ref{gpu:verts_begin} where we add the truth value of a logical expression, even if it evaluates to 0.%(On the other hand we do allow branching within this logical expression, expecting the cost of accessing the voxel value to be greater than that incurred by branching.)

\myparagraph{Constant memory.} Kernels often require additional information, e.g. the width and height of the processed image. Accessing such information from global memory multiple times would be inefficient. Instead, we use ensure quick access by declaring such variables as constant memory, see e.g. line~\ref{gpu:const}. With this, the values are stored in GPU's \emph{constant memory}, which is specialized for fast \emph{broadcast} of stored values to multiple threads.

\begin{figure*}[!t]
\centering
\includegraphics[width=.85\textwidth]{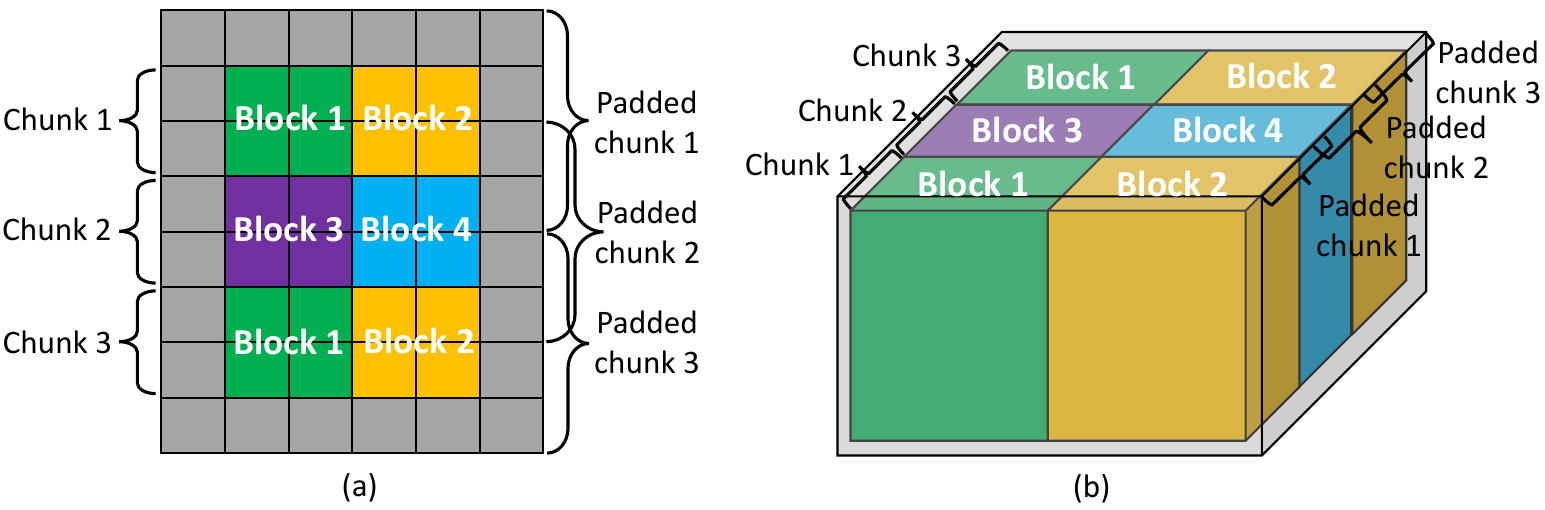}
\caption[ph]{Chunking in 2D and 3D. We emphasize that the chunks are a disjoint decomposition of the input -- but the padded chunks are not. This extra padding provides information necessary to ensure that each cell introduced by input voxels is counted exactly once.}
\label{fig:streaming}
\end{figure*}

\subsection{Streaming}
\label{sec:streaming}
In practice, a lot of 3D images are too large to fit in GPU memory. We overcome this obstacle by dividing an input into \emph{chunks} and process them separately. Another benefit of streaming lies in the CUDA's asynchronous behaviour which allows us to overlap data transfers and kernel executions. Breaking an input into smaller pieces helps hide the high latency related to memory transfers between GPU and RAM; see Fig.~\ref{fig:sync_connect}.

Input of size $(w_0, w_1,\dots)$ is cut along the first coordinate. This way the image is divided into $c$ \emph{chunks} of size at most $(\floor*{\frac{w_0}{c}}, w1, \dots)$. This ensures that the resulting chunks correspond to contiguous memory addresses as arrays are stored in row-major order in C++. As illustrated in Fig. \ref{fig:streaming}, we extend the chunks by a single-voxel padding. The collar contains either the value of a voxel -- to ensure that each input voxel has access to all of its neighbours; or positive infinity -- as explained before.

Each chunk is loaded into the GPU memory (including the collar). The chunk is then processed with one or multiple CUDA blocks depending on the size. After finishing computations, the free blocks will be reassigned to a new chunk for computation.

\myparagraph{Overlapping computations and data transfers.} CUDA devices contain engines for various tasks. Modern devices typically have two copy engines, one for host-to-device transfers and another for device-to-host transfers, as well as a kernel engine. With pinned (non-pageable) host memory, the tasks launched into non-default different CUDA streams can be executed concurrently assuming no dependencies amongst them. In other words, loading a chunk into device, writing results back to host, and kernel execution can happen simultaneously. With a reasonable choice of $c$, the overhead of data transfer can be greatly alleviated. Fig. \ref{fig:sync_connect}(b) illustrates a simplified case of overlapping transfers and kernel executions. Suppose we have equal running time for memory copy and kernel execution. Compared to serialized execution, overlapped execution practically hides the kernel execution time for one chunk when $c = 2$.

%\myparagraph{Automatic mechanisms.} Modern GPUs offer additional mechanisms which aim at helping the programmer. In particular, similarly to CPUs, GPUs offer multiple levels of automatic cache memory. Also the GPU schedules individual warps specifically to hide the high memory-access latency. Still, compared to CPU programming an efficient GPU implementation requires the programmer to explicitly control many types of memory, as well as other hardware-specific factors.

% ============ Major Performance =====
\begin{table}[t]
\small
\begin{center}
\caption{This table compares the execution time of the CPU and GPU implementations. These are end-to-end timings, include disk I/O and the GPU overhead related to initializing our computations. The two rightmost columns are relevant in situations in which  the input resides in GPU memory.}
\label{table:overall}
\ra{0.6}
\begin{tabular}{@{}r | P{0.8cm} P{0.8cm} P{0.8cm} P{0.8cm} P{0.8cm} P{0.8cm} P{0.8cm} | P{0.8cm} P{0.8cm}@{}}
\toprule
\multicolumn{1}{r}{} 
& \multicolumn{1}{c}{} 
& \multicolumn{1}{c}{} 
& \multicolumn{1}{c}{} 
& \multicolumn{1}{c}{} 
& \multicolumn{1}{c}{\textBF{CPU}}
& \multicolumn{1}{c}{\textBF{GPU}}
& \multicolumn{1}{c}{\textBF{GPU}}
& \multicolumn{1}{c}{\textBF{GPU}} 
& \multicolumn{1}{c}{\textBF{GPU}} \\

\multicolumn{1}{r}{} 
& \multicolumn{1}{c}{\textBF{Input}} 
& \multicolumn{1}{c}{\textBF{CPU}} 
& \multicolumn{1}{c}{\textBF{GPU}} 
& \multicolumn{1}{c}{\textBF{Overall}} 
& \multicolumn{1}{c}{\textBF{disk}}
& \multicolumn{1}{c}{\textBF{disk}} 
& \multicolumn{1}{c}{\textBF{over-}} 
& \multicolumn{1}{c}{\textBF{exec.}} 
& \multicolumn{1}{c}{\textBF{kernel}} \\

\multicolumn{1}{r}{} 
& \multicolumn{1}{c}{\textBF{size(B)}} 
& \multicolumn{1}{c}{\textBF{overall}} 
& \multicolumn{1}{c}{\textBF{overall}} 
& \multicolumn{1}{c}{\textBF{speedup}} 
& \multicolumn{1}{c}{\textBF{read}}
& \multicolumn{1}{c}{\textBF{read}} 
& \multicolumn{1}{c}{\textBF{head}} 
& \multicolumn{1}{c}{\textBF{(kernel)}} 
& \multicolumn{1}{c}{\textBF{Gvox/s}} \\

\cmidrule{2-10}
\multicolumn{1}{r}{} & \multicolumn{9}{c}{Uniform Noise}\\
\midrule
4096$^3$ & 256G & 37.72m & 9.10m & 4.14x & 7.30m & 9.08m & 0.67s & 0.20m & 5.62\\
2048$^3$ & 32G  & 4.86m  & 0.71m & 6.77x & 0.99m & 0.71m & 0.41s & 0.03m & 5.61\\
1024$^3$ & 4G   & 36.85s & 5.63s & 6.55x & 6.85s & 5.20s & 0.37s & 0.16s & 6.57\\
512$^3$  & 512M & 4.97s  & 0.85s & 5.86x & 1.00s & 0.64s & 0.19s & 0.02s & 6.55\\

\midrule
\multicolumn{1}{r}{} & \multicolumn{9}{c}{Gaussian Random Field}\\ \midrule
512$^3$  & 512M & 4.93s & 0.86s & 5.75x & 0.90s & 0.66s & 0.19s & 20.88ms & 6.43\\
256$^3$  & 64M  & 0.63s & 0.24s & 2.58x & 0.13s & 0.09s & 0.15s & 2.64ms  & 6.35\\
128$^3$  & 8M   & 0.11s & 0.12s & 0.86x & 0.02s & 0.01s & 0.12s & 0.35ms  & 6.03\\
8192$^2$ & 256M & 1.47s & 0.53s & 2.75x & 0.44s & 0.36s & 0.16s & 6.64ms  & 10.10\\
4096$^2$ & 64M  & 0.38s & 0.21s & 1.84x & 0.12s & 0.08s & 0.14s & 1.74ms  & 9.67\\
2048$^2$ & 16M  & 0.09s & 0.18s & 0.55x & 0.04s & 0.03s & 0.12s & 0.45ms  & 9.36\\

\midrule
\multicolumn{1}{r}{} & \multicolumn{9}{c}{VICTRE}\\ \midrule
287 359 202 & 79.3M & 0.59s & 0.30s & 1.98x & 0.16s & 0.13s & 0.14s & 3.85ms  & 5.41\\
440 518 488 & 424M  & 2.99s & 0.77s & 3.87x & 0.98s & 0.45s & 0.24s & 20.65ms & 5.39\\
434 446 384 & 147M  & 1.11s & 0.36s & 3.02x & 0.29s & 0.15s & 0.16s & 7.13ms  & 5.40\\
434 446 384 & 283M  & 1.96s & 0.53s & 3.70x & 0.79s & 0.30s & 0.18s & 13.72ms & 5.42\\
 
\midrule
\multicolumn{1}{r}{} & \multicolumn{9}{c}{CMB}\\ \midrule
1500 750  & 1.07M & 0.03s & 0.12s & 0.22x & 0.01s & 0.01s & 0.11s & 0.15ms & 7.40\\
3000 1500 & 4.29M & 0.09s & 0.15s & 0.61x & 0.04s & 0.02s & 0.13s & 0.44ms & 10.16\\
6400 3200 & 19.5M & 0.37s & 0.25s & 1.49x & 0.13s & 0.08s & 0.14s & 1.94ms & 10.56\\

\bottomrule
\end{tabular}
\end{center}
\end{table}

\section{Experiments}
\label{sec:experiment}
We use the C++ compiler shipped with Visual Studio 2019 (v142) and language standard of C++14 for the compilations of both CPU and GPU implementations. The following experiments are conducted on a desktop machine with Intel Core i7-9700K CPU with 8 physical cores (and disabled hyper-threading), 16GB of RAM, Sabrent Rocket Q 2TB NVMe PCIe M.2 2280 SSD drive, and a NVIDIA RTX 2070 graphics card with 8GB of GDDR6 memory. It is a modern commodity workstation.

\myparagraph{Datasets.}
We use a mix of synthetic and real-world datasets:

\begin{itemize}
    \item Cosmic microwave background (CMB) imaging data comes from astrophysical measurements. The original data is on a 2-dimensional sphere; we use a single image projection in different resolution. Each image contains at most 256 unique values.
    
    \item Virtual Imaging Clinical Trials for Regulatory Evaluation (VICTRE)~\cite{VICTRE01} project provides realistic simulation of breast phantoms. We generated 20 3D breast volumes. Each image contains only 11 unique values.

    \item We also use a set of 70 2D Gaussian Random Fields (GRF) with 7 sizes (10 samples for each size) and 30 3D GRFs with 3 sizes (10 samples each). Each image contains only 1024 unique values.

    \item For larger experiments we use data generated by sampling the uniform distribution for each voxel. We call this data uniform noise.
\end{itemize}

All datasets except for CMB are stored in binary format as 32 bit IEEE 754 floating point values. CMB is stored in binary format as 8-bit unsigned integer values.

\myparagraph{Voxel throughput.} We are mostly interested in the size (counted in numbers of pixels or voxels) of the image that can be processed in a second. We call this quantity the \emph{voxel throughput} and express it in GVox/s, namely billions ($10^9$) voxels per second. All time measurements are given in ms (milliseconds, $10^{-3}$s).

% ============ Repeated Gaussian =====
\begin{table}[t]
\small
\begin{center}
\caption{This table shows the timings for the pipeline involving the iterated ECC and Gaussian smoothing computations. The key observations is that when averaged over multiple iterations the overall time is dominated by the two kernel executions. This confirms that there are no additional bottlenecks in this pipeline, and especially in our ECC computations. Note that the image is read once, and so the time to load the image from disk is a one-time cost.}
\label{table:gaussian}
\ra{0.95}
\begin{tabular}{@{}r | P{0.8cm} P{0.8cm} P{0.8cm} P{0.8cm} P{0.8cm} P{0.8cm}@{}}
\toprule
\multicolumn{1}{r}{} 
& \multicolumn{1}{c}{} 
& \multicolumn{1}{c}{\textBF{Overall}} 
& \multicolumn{1}{c}{\textBF{ECC mem.}} 
& \multicolumn{1}{c}{\textBF{ECC exec.}} 
& \multicolumn{1}{c}{\textBF{Gaussian}}
& \multicolumn{1}{c}{\textBF{Disk}} \\

\multicolumn{1}{r}{} 
& \multicolumn{1}{c}{\textBF{Overall}} 
& \multicolumn{1}{c}{\textBF{avg.}} 
& \multicolumn{1}{c}{\textBF{avg.}} 
& \multicolumn{1}{c}{\textBF{avg.}} 
& \multicolumn{1}{c}{\textBF{exec. avg.}}
& \multicolumn{1}{c}{\textBF{read}} \\

\multicolumn{1}{r}{} 
& \multicolumn{1}{c}{\textBF{[ms]}} 
& \multicolumn{1}{c}{\textBF{[ms]}} 
& \multicolumn{1}{c}{\textBF{[ms]}} 
& \multicolumn{1}{c}{\textBF{[ms]}} 
& \multicolumn{1}{c}{\textBF{[ms]}}
& \multicolumn{1}{c}{\textBF{[ms]}} \\

\cmidrule{2-7}
\multicolumn{1}{r}{} & \multicolumn{6}{c}{Uniform Noise}\\
\midrule
(ECC+Gaussian) $\times$ 1    & 137.16  & 137.16 & 0.28 & 0.16 & 1.55 & 7.72\\
(ECC+Gaussian) $\times$ 10   & 172.80  & 17.28  & 0.06 & 0.15 & 0.20 & 7.38\\
(ECC+Gaussian) $\times$ 100  & 149.96  & 1.50   & 0.03 & 0.13 & 0.09 & 7.81\\
(ECC+Gaussian) $\times$ 1000 & 352.02  & 0.35   & 0.03 & 0.12 & 0.07 & 7.22\\
(ECC+Gaussian) $\times$ 1000 & 2786.64 & 0.28   & 0.03 & 0.17 & 0.07 & 7.57\\

\bottomrule
\end{tabular}
\end{center}
\end{table}

\subsection{Case study: Single image on disk}
In this case we employ CHUNKYEuler by Heiss and Wagner \cite{chunkyeuler} as a CPU baseline. CHUNKYEuler is the state-of-the-art CPU parallel streaming ECC implementation. To the best of our knowledge, no other software can handle the sizes of the data we experiment with. We run experiments with all eight available CPU cores.

\myparagraph{Overall execution time.} 
In this setup, we simply measure the overall execution time including reading the image from disk; see Table~\ref{table:overall}. For files smaller than around 16MB, the CPU version is faster. This is due to the overhead related to initializing our GPU computations. For files larger than 0.5GB, the GPU version is between 4 to 6 times faster -- although it is severely limited by disk I/O which takes between $75\%$ and $99.7\%$ of its total execution time.

\myparagraph{Streaming.} Note that we handle files significantly larger than the available 8GB GPU memory and 16GB RAM. This is achieved by a streaming algorithm described before. This was a major difficulty and is described in Section~\ref{fig:streaming}. In particular, we handled an image of size $4096^3$ which takes $0.25$TB.

\myparagraph{GPU overhead.} The overhead mentioned above is related to the initialization and shutdown of the GPU device, and memory allocation specific to our implementation. This overhead ranges between 100 and 700ms and is a one-time cost. This is why GPU is more effective for larger datasets -- but also for batches of smaller ones. We will focus on that next.

\subsection{Case study: Batch processing of images on disk}
In this case we read multiple files from disk. We focus on small files, because they were problematic for the GPU implementation (due to the GPU overhead). Table~\ref{table:batch} shows that the overhead now amortizes when many files are processed. This means that in batch processing the GPU implementation is always preferred over the CPU one. Still, this is not an ideal setup for GPU, since the computations are heavily limited by disk I/O.

\myparagraph{Prospects.} The above issue opens up a new avenue -- it may now be opportune to load compressed images, which would limit the disk I/O time. We plan to investigate this in future work.

\subsection{Case study: GPU-only pipeline}
In this scenario, the images are stored and processed entirely in GPU memory. This emulates pipelines implemented entirely on GPUs, such as some implementations of CNNs~\cite{li2016performance}. As mentioned earlier, this case is our primary motivation. We are trying to determine if our ECC kernel could be part of such a GPU pipeline without becoming a significant performance bottleneck. We also need to verify that our computational setup does not incur any unexpected additional bottlenecks.

% ============ Batch Processing =====
\begin{table}[t]
\small
\begin{center}
\caption{We show timings averaged over running different numbers of files. This table confirms that the GPU overhead, which dominates the computations for a single small file, amortizes across many samples. It is clear that the GPU performance is heavily limited by disk I/O.}
\label{table:batch}
\ra{0.95}
\begin{tabular}{@{}r | P{0.8cm} P{0.8cm} P{0.8cm}@{}}
\toprule
\multicolumn{1}{r}{} 
& \multicolumn{1}{c}{Input} 
& \multicolumn{1}{c}{GPU} 
& \multicolumn{1}{c}{GPU disk}\\

\multicolumn{1}{r}{} 
& \multicolumn{1}{c}{\textBF{size(B)}} 
& \multicolumn{1}{c}{\textBF{overall avg. [ms]}} 
& \multicolumn{1}{c}{\textBF{read avg. [ms]}}\\

\cmidrule{2-4}
\multicolumn{1}{r}{} & \multicolumn{3}{c}{Uniform Noise}\\
\midrule
128$^2 \times$ 1     & 64K   & 119.83 & 0.69\\
128$^2 \times$ 100   & 6.25M & 1.77   & 0.46\\
128$^2 \times$ 1000  & 62.5M & 0.66   & 0.45\\
128$^2 \times$ 10000 & 625M  & 0.52   & 0.42\\

\midrule
\multicolumn{1}{r}{} & \multicolumn{3}{c}{Gaussian Random Field}\\ \midrule
128$^3 \times$ 1    & 8M    & 124.68 & 12.02\\
128$^3 \times$ 10   & 80M   & 28.13  & 13.86\\
128$^3 \times$ 100  & 800M  & 15.38  & 13.82\\
128$^3 \times$ 1000 & 8000M & 11.96  & 11.67\\

\bottomrule
\end{tabular}
\end{center}
\end{table}

\myparagraph{Pipeline.} To this aim, we consider a two-step pipeline: (1) compute the ECC; (2) apply a Gaussian smoothing filter. Steps (1) and (2) are performed repeatedly on an image stored in GPU memory. We iterate up to 10000 times using a $1024^2$ GRF image. After each iteration, the resulting VCEC is transferred to RAM and post-processed, including computing ECC.

\begin{figure}[ht]
    \centering
    \includegraphics[width=1.\textwidth]{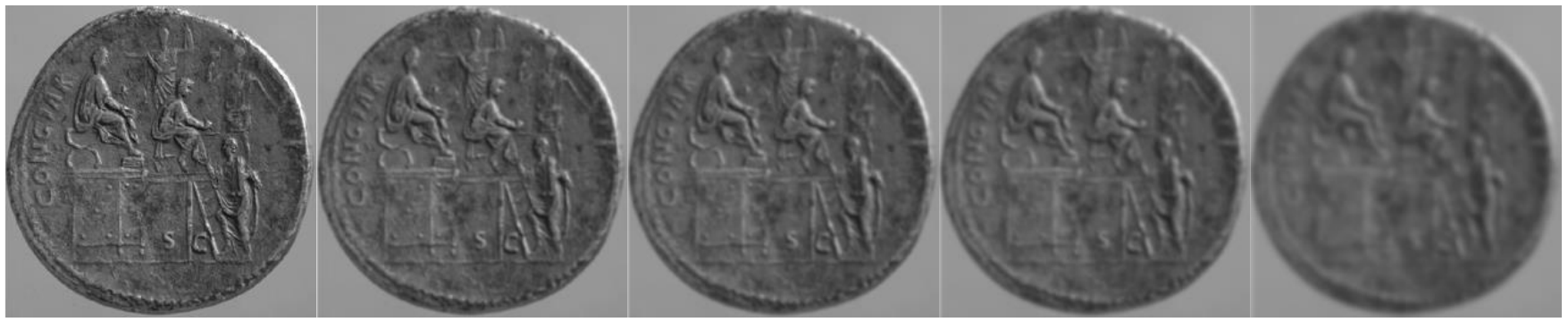}
    \caption{Images at consecutive steps in the smoothing pipeline. %Note that smoothing mitigates the main issue with ECC, namely its %tendency to pick up noisy topological features (as already %illustrated in Figure~\ref{fig:pers_to_euler}).
    }
    \label{fig:my_label}
\end{figure}

\myparagraph{Gaussian smoothing implementation.} We implement the Gaussian smoothing filter as a discrete Gaussian convolution. We exploit its separability and use a highly optimized GPU kernel. We use a Gaussian kernel width of 13 pixels (see also~Fig. \ref{fig:my_label}).

%\myskip{
%}

\myparagraph{Potential performance bottlenecks.} Since the initial image is loaded into GPU memory once, the cost of reading from disk amortizes across many kernel runs. As we already checked, the same applies to the GPU overhead. Column "ECC mem" in Table~\ref{table:gaussian} shows the cost of transferring the resulting VCEC from GPU memory to RAM and the cost of its CPU post-processing; this does not incur a performance hit either. Overall, we see that the kernel executions dominate the overall time. 

\myparagraph{Performance comparison.} We can therefore directly compare the performance of the ECC kernel and the convolution kernel.  Table~\ref{table:gaussian} shows that the throughput of the two kernels is at the same order of magnitude. The Gaussian kernel is up to 2.5 times faster. However, the impact on the overall performance of a CNN is likely to be significantly lower, since a single convolution often contributes less than half of the total computation time performed by a convolution layer in a CNN~\cite{li2016performance}.

\myparagraph{ECC kernel performance.} We highlight the performance of the ECC kernel. The throughput is between 5 and 10 GVox/s. To put things in perspective, it allows us to handle:
\begin{enumerate}
    \item 3D images of size $512^3$ voxels at the rate of 30Hz;
    \item 2D images of 8K resolution ($7680 \times 4320$ pixels) at the rate of 120Hz.
\end{enumerate}

\subsection{Dependence on dimension} 
Perhaps surprisingly, the performance does not depend on the dimension of the image -- which suggests that the caching hierarchy we devised works well --  and that the neighbours are typically retrieved from the cache. This way the dependence on the number of neighbours (8 vs 26) largely disappears. This property would not extend to higher dimensions, since the texture cache is only available in dimensions that are smaller or equal to 3.

\section{Discussion}
We proposed an efficient GPU implementation to compute the Euler characteristic curve of imaging data. The resulting software is highly practical. Its three major advantages are:

\begin{itemize}
    \item High speed: for images present in GPU memory, it processes images at speed exceeding $5\times 10^9$ voxels per second. This is a realistic scenario for example in the context of convolutional networks.
    \item Streaming: it can handle images of virtually unlimited size. This is crucial since GPU memory is a limited resource.
    \item ECC contains topological information which was successfully used in many application domains.
\end{itemize}

We believe these results open up interesting avenues. Our plans are twofold. First, we intend to integrate our ECC computations into CNNs. With the efficiency gap closed, we hope that topological methods will start permeating  mainstream machine learning. Second, we hope that the full power of persistent homology can be used in such contexts. With the gathered experience specific to handling cubical filtrations on GPUs, we hope to make the first steps towards designing GPU algorithms for persistence analysis of imaging data.

\bibliographystyle{plainurl}
\bibliography{main}

\end{document}